\documentclass[12pt]{article}
\usepackage{latexsym}
\usepackage{graphicx}
\usepackage[english]{babel}
\begin{document}
\input epsf

\def\p{\partial}
\def\h{{1\over 2}}
\def\be{\begin{equation}}
\def\bea{\begin{eqnarray}}
\def\ee{\end{equation}}
\def\eea{\end{eqnarray}}
\def\d{\partial}
\def\la{\lambda}
\def\eps{\epsilon}
\def\bb{\bigskip}
\def\mm{\medskip}
\newcommand{\dm}{\begin{displaymath}}
\newcommand{\edm}{\end{displaymath}}
\renewcommand{\b}{\tilde{B}}
\newcommand{\gm}{\Gamma}
\newcommand{\ac}[2]{\ensuremath{\{ #1, #2 \}}}
\renewcommand{\ell}{l}
\newcommand{\z}{\ell}
\newcommand{\newsection}[1]{\section{#1} \setcounter{equation}{0}}
\def\bb{$\bullet$}
\def\Qbar{{\bar Q}_1}
\def\QPbar{{\bar Q}_p}

\def\q{\quad}

\def\bn{B_\circ}

\let\a=\alpha \let\b=\beta \let\g=\gamma \let\d=\delta \let\e=\epsilon
\let\c=\chi \let\th=\theta  \let\k=\kappa
\let\l=\lambda \let\m=\mu \let\n=\nu \let\x=\xi \let\r=\rho
\let\s=\sigma \let\t=\tau
\let\vp=\varphi \let\vep=\varepsilon
\let\w=\omega      \let\G=\Gamma \let\D=\Delta \let\Th=\Theta
                     \let\P=\Pi \let\S=\Sigma

\def\h{{1\over 2}}
\def\t{\tilde}
\def\r{\rightarrow}
\def\nn{\nonumber\\}
\let\bm=\bibitem
\def\Kt{{\tilde K}}
\def\b{\bigskip}

\let\p=\partial

\begin{flushright}
\end{flushright}
\vspace{20mm}
\begin{center}
{\LARGE  Membrane paradigm realized?\footnote{(Slightly expanded version)  of essay written for the Gravity Research Foundation 2010 essay contest.}}
\\
\vspace{18mm}
{\bf  Samir D. Mathur }\\

\vspace{8mm}
Department of Physics,\\ The Ohio State University,\\ Columbus,
OH 43210, USA\\ \vskip .2 in   mathur@mps.ohio-state.edu
\vspace{4mm}
\end{center}
\vspace{10mm}
\thispagestyle{empty}
\begin{abstract}

\b

Are there any degrees of freedom on the black hole horizon?  Using the `membrane paradigm'  we can reproduce coarse-grained physics outside the hole by assuming a fictitious membrane just outside the horizon.  But to solve the information puzzle we need  `real' degrees of freedom at the horizon, which can  modify Hawking's evolution of  quantum modes. We argue that recent results on gravitational microstates imply a set of real degrees of freedom just outside the horizon; the state of the hole is a linear combination of rapidly oscillating gravitational solutions  with support concentrated just outside the horizon radius. The collective behavior of these microstate solutions may give a realization of the membrane paradigm, with the fictitious membrane now replaced by  real, explicit degrees of freedom.

\end{abstract}
\vskip 1.0 true in

\newpage
\setcounter{page}{1}

A longstanding puzzle in quantum gravity is the following: is there anything at the black hole horizon?

There are many reasons to take the horizon seriously as a place where `physics happens'. 
The Bekenstein entropy $S_{bek}$ of the hole is proportional to the area of the horizon \cite{bek}. This suggests that we attach one `bit' of information to each planck area of the horizon, and perhaps quantize the horizon area to give an integral number of bits \cite{quant}. In loop quantum gravity the entropy arises from spin degrees of freedom on links that intersect the horizon \cite{loop}. In the Schwarzschild coordinate frame, all matter that ever fell towards the horizon appears to be frozen forever just outside the horizon. This picture was concretized in the `membrane paradigm' \cite{membrane}, where one puts an artificial cutoff surface just outside the horizon, and finds boundary conditions at this surface which will reproduce correct mechanical and thermal  effects in the region  outside this surface. These boundary conditions can be summarized by imagining a fictitious membrane placed at this surface, with appropriate mechanical and thermal properties. In particular, there is a very high temperature near the membrane, so one might imagine that an infalling observer would get burnt up and eventually reradiated out to infinity \cite{susskind}. This would be good, because to avoid the information paradox we need to have some way of returning the information in infalling objects back to infinity.

But there turns out to be a flaw in these attempts to assign genuine dynamics to the horizon.  It is well known that the Schwarzschild coordinate system fails at the horizon, and the regular coordinates there -- the Kruskal coordinates -- show that {\it nothing} special happens at the horizon of the traditional black hole. Nothing can `stay' at the horizon; all matter must  fall in towards the singularity. The high temperature in the Schwarzschild frame disappears in the Kruskal frame; it is seen to be nothing but the Rindler tempertaure observed in an accelerated system of coordinates. Careful renormalization of the stress tensor shows that there is no energy density at the horizon  \cite{many};  the energy of Rindler quanta is cancelled by vacuum polarization energy, since the physical vacuum at the horizon is naturally described in Kruskal rather than Schwarzschild coordinates.

What then, about the fate of information falling into the hole? Over the past decade, many physicists had settled down to  the following set of beliefs (i) information is not lost; it emerges in the Hawking radiation  from the hole (ii) nothing special happens at the horizon, because the Kruskal description looks like the logical one  (iii) the above two beliefs are consistent because we need only small delicate correlations among Hawking quanta to carry the information, and subleading corrections to Hawking's computation \cite{hawking} should be able to encode the required information in the quanta.

But if these beliefs were correct, one has to wonder why over the years no one produced a model showing how it was possible for information to get encoded in the radiation while preserving a near-semiclassical Hamiltonian at the horizon. The reason for this failure emerged last year, when  it was proved that (iii) is in fact incorrect: it is {\it not} possible to make small corrections to Hawking's leading order state and recover any significant information in the radiation \cite{mathur1}. More precisely, if matrix elements of the Hamiltonian for low energy modes are within $\epsilon$ of their semiclassical values, then no more than a fraction $2\epsilon$ of the information can be encoded in the radiation. The proof of this theorem used a well known inequality from quantum information theory called `strong subadditivity' \cite{lieb}. But no {\it elementary} proof of this inequality is known, so it is perhaps not surprising that its consequences were not built into our naive thinking about encoding of information in the radiation. 

It would now seem that we are back to square one on the information puzzle. Must we reconcile ourselves to information loss? We will now argue that another set of results over the past few years gives us a picture of how there can be real dynamics at the horizon, which will lead to an escape of information in the radiation.

A black hole has  given values of of its conserved quantum numbers: mass, charge and angular momentum.  In earlier work, one assumed a simple ansatz for the metric. Thus for the nonrotating hole, one assumed  independence of the angular directions $\theta, \phi$ for all metric components.  In a higher dimensional gravity theory, the extra directions have to be assumed compact, and the ansatz assumed no dependence of the metric on the compact coordinates $z_i$ either. Einstein's equations then gave a unique solution, possessing a horizon and a singularity -- the traditional black hole.

 But it has been now recognized that there are $Exp[S_{bek}]$ {\it different} solutions of  Einstein's equations with the {\it same} quantum numbers \cite{lm4,fuzzballs}. All are regular, with no horizon and no singularity. None of them are spherically symmetric, and all involve the compact directions in a nontrivial way. So it was the simplicity of our initial ansatz which led us to miss these solutions. Further, string theory has allowed us to list all possible states for simple sets of quantum numbers.  The regular states mentioned above map in a one-to one and onto way onto the set of all allowed states, while the spherical solution with horizon and singularity is not realized at all. 
 
 This situation gives us an appealing way out of the information problem:
 
 \b
 
\begin{figure}[htbp]
\begin{center}
\includegraphics[scale=.15]{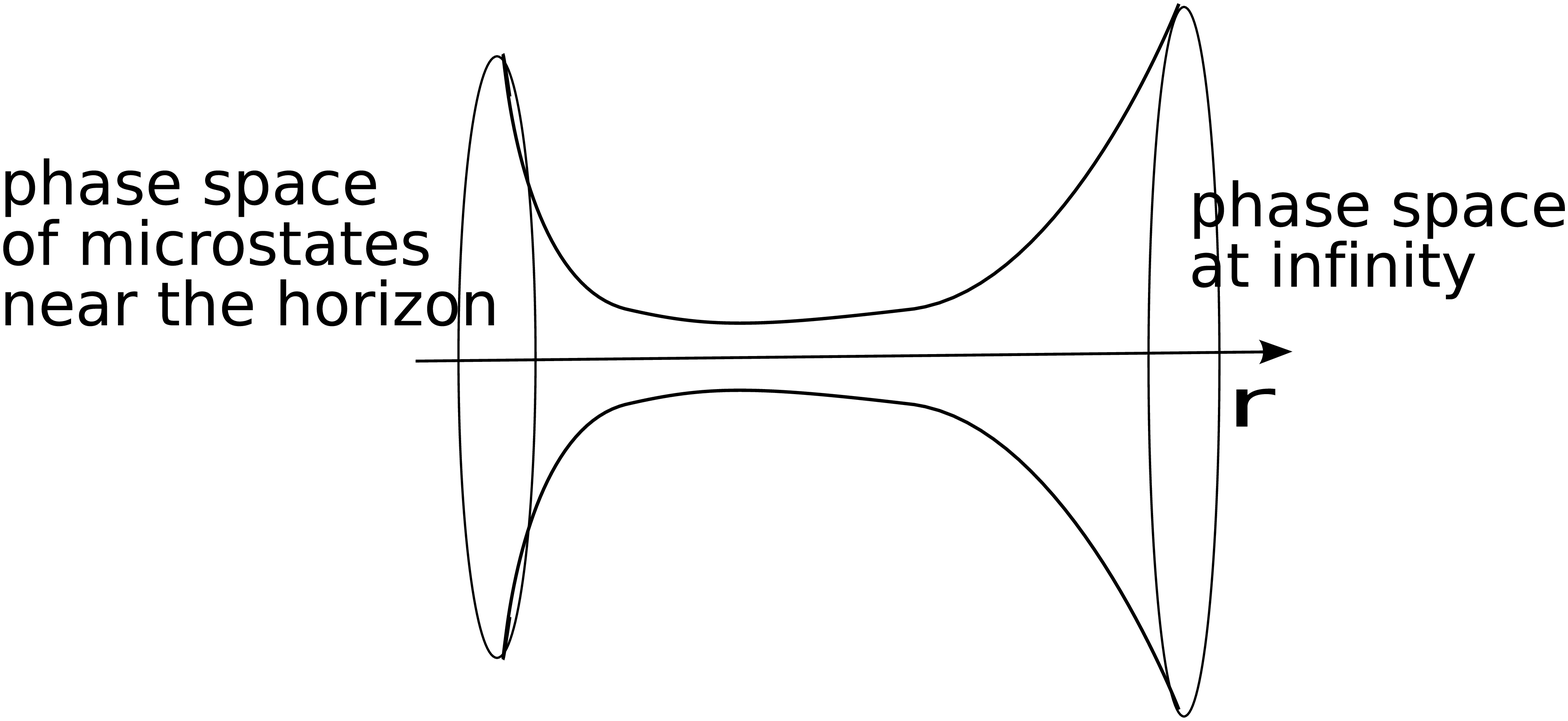}
\caption{Schematic depiction of phase space in the black hole: there is a large phase space near the horizon from the microstate solutions of Einstein's equations.}
\label{fone}
\end{center}
\end{figure}

 (a) The region far from the hole has an infinite dimensional phase space due to its infinite volume. Very close to the horizon, however, there is again a very large phase space \cite{phase}, from the $Exp[S_{bek}]$ solutions -- termed 'microstates' -- giving the entropy of the hole (fig.\ref{fone}). 
 
 \b
 
\begin{figure}[htbp]
\begin{center}
\includegraphics[scale=.15]{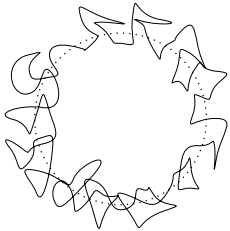}
\caption{Schematic description of a microstate solution of Einstein's equations. There are `local ergoregions' with rapidly changing direction of frame dragging near the horizon. The geometry closes off without having an interior horizon or singularity due to its peculiar topological structure.}
\label{ftwo}
\end{center}
\end{figure}

\begin{figure}[htbp]
\begin{center}
\includegraphics[scale=.15]{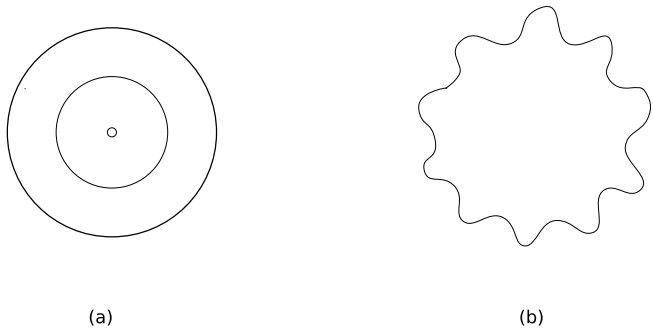}
\caption{A circular elastic band in flat space will shrink to a point under its tension, but rapid oscillations maintain a nonzero average radius for the loop.}
\label{fthree}
\end{center}
\end{figure}

 (b) Most of these solutions have nontrivial structure only very close to the horizon. This is because rapid variation of the metric with position costs a lot of energy, and the large redshift near $r=2M$ is needed to lower this to the given total energy $M$. The detailed structure of the solutions can be inferred from the explicitly solutions which have been constructed. The geometry has {\it ergoregions}, with the direction of frame dragging changing rapidly from place to place (fig.\ref{ftwo}). This rapid space-time time dependence is what keeps the geometrical structure from collapsing towards $r=0$. (For an analogy, see fig.\ref{fthree}; a circular  elastic band shrinks to a point, but an rapidly vibrating band stays open with a nontrivial average radius.)
 
 \b
 
 (c) Consider a spherically symmetric shell made of gravitational radiation, collapsing to make a hole. The motion upto $r= 2M+\epsilon$ is described by traditional dynamics in the Schwarzschild frame ($\epsilon\ll 2M$). But at $r<2M+\epsilon$ the wavefunction of the shell can spread over the $Exp[S_{bek}]$ alternative solutions of Einstein's equations. The amplitude for this spread can be estimated by computing the {\it tunneling} amplitude between the shell and one of the microstate solutions. This amplitude $Exp[-S_{cl}]$ is very {\it small}, since the involved objects are macroscopic and the classical action of tunneling $S_{cl}$ is big.  But the {\it number} of  states that we can tunnel to very {\it large}, given by $Exp[S_{bek}]$. It turns out that $S_{cl}\sim S_{bek}$ \cite{tunnel}, so instead of making a long lived black hole, the shell state becomes a linear superposition of microstate solutions in a time which is much shorter than the Hawking evaporation time \cite{time}.

 \b
 
 (d) It has been checked that ergoregion emission from the microstate solutions reproduces exactly the Hawking radiation rate expected from those solutions \cite{radiation}. But this emission does not involve information loss, since there is no horizon. 
 
 \b
 
 (e) The further evolution of the collapsing shell  is given by the evolution of the linear superposition of microstates into which it has tunneled.   It is plausible that the short time evolution of the collective modes of these complicated microstates can be mimicked by evolution in the interior of the traditional black hole geometry. This effective evolution would then corresponds to the interior of the black hole in the Penrose diagram. It is important to note here that the typical microstate is not a classical solution in which infalling quanta scatter around in a zig-zag path. Instead, the large phase space of fuzzball solutions  creates a very large `coordination number' -- the number of states that can be reached from a  given state by a small action $\Delta S$. (This may make the black hole a `fast scrambler' \cite{scrambler}.) An infalling heavy object  excites a linear combination of fuzzball states, similar to the `fermi-golden rule' absorption encountered when we have a closely spaced band of energy levels.  Different objects excite different degrees of freedom to leading order, since the number of excitable states is so large; the small overlap between different sets of excited states gives the residual interaction that mimics the usual (weak) gravitational interaction between different quanta (the dual CFT description of this effect was noted in \cite{lm4}). Thus the effective dynamics of  absorption into fuzzballs gets related to motion in gently curved space, in a manner similar to absorption by the large number of open string degrees of freedom on a stack of D-branes: the infalling quantum excites these open strings, different quanta typically excite different strings (due to the large number of these strings), and the small overlap between the set of excited strings generates the effective gravitational interactions between infalling quanta. Thus in this picture we would think of both AdS/CFT duality  \cite{maldacena} and the above mentioned effective description of fuzzballs as arising from the same underlying physics: absorption into a dense set of levels with high coordination number.
 
 \b
 
 With this picture we seem to have an explicit realization of the membrane paradigm. In several earlier works, effective gravitational dynamics or  dual CFT descriptions have been used to infer an imaginary `membrane' that may sit at the horizon \cite{earlier}. By contrast, we are now addressing  how gravity manages to actually produce something at the horizon that is capable of sending information back out in the Hawking radiation. In our description the membrane  has real degrees of freedom on it, given by the rapidly oscillating solutions of Einstein's equations that  correspond to the black hole microstates.  It would be exciting to explore if the collective dynamics of the microstates reproduces the very precise properties of the fictitious membrane of the membrane paradigm.
 
\section*{Acknowledgements}

This work was supported in part by DOE grant DE-FG02-91ER-40690.


\newpage

\begin{thebibliography}{99}

 \bibitem{bek}
J.~D.~Bekenstein,
Phys.\ Rev.\ D {\bf 7}, 2333 (1973).
%

\bibitem{quant}
  J.~D.~Bekenstein,
  Lett.\ Nuovo Cim.\  {\bf 11}, 467 (1974).

\bibitem{loop}
  C.~Rovelli and L.~Smolin,
  Nucl.\ Phys.\  B {\bf 442}, 593 (1995)
  [Erratum-ibid.\  B {\bf 456}, 753 (1995)]
  [arXiv:gr-qc/9411005].
  
  \bibitem{membrane}
  K.~S.~.~Thorne, R.~H.~.~Price and D.~A.~.~Macdonald,
{\it  NEW HAVEN, USA: YALE UNIV. PR. (1986) 367p}

\bibitem{susskind}
  L.~Susskind, L.~Thorlacius and J.~Uglum,
  Phys.\ Rev.\  D {\bf 48}, 3743 (1993)
  [arXiv:hep-th/9306069].


\bibitem{many}
  D.~W.~Sciama, P.~Candelas and D.~Deutsch,
  Adv.\ Phys.\  {\bf 30}, 327 (1981).

  
 
  \bibitem{hawking}
  S.~W.~Hawking,
  Commun.\ Math.\ Phys.\  {\bf 43}, 199 (1975)
  [Erratum-ibid.\  {\bf 46}, 206 (1976)];
  S.~W.~Hawking,
  Phys.\ Rev.\  D {\bf 14}, 2460 (1976).


\bibitem{mathur1}
  S.~D.~Mathur,
  Class.\ Quant.\ Grav.\  {\bf 26}, 224001 (2009)
  [arXiv:0909.1038 [hep-th]].

\bibitem{lieb}
  E.~H.~Lieb and M.~B.~Ruskai,
  J.\ Math.\ Phys.\  {\bf 14}, 1938 (1973).


  
 \bibitem{lm4}
 O.~Lunin and S.~D.~Mathur,
  Nucl.\ Phys.\  B {\bf 623}, 342 (2002)
  [arXiv:hep-th/0109154];


 \bibitem{fuzzballs}
     O.~Lunin and S.~D.~Mathur,
  Phys.\ Rev.\ Lett.\  {\bf 88}, 211303 (2002)
  [arXiv:hep-th/0202072];
  O.~Lunin, J.~M.~Maldacena and L.~Maoz,
  arXiv:hep-th/0212210;
I.~Kanitscheider, K.~Skenderis and M.~Taylor,
  JHEP {\bf 0706}, 056 (2007)
  [arXiv:0704.0690 [hep-th]].
S.~Giusto, S.~D.~Mathur and A.~Saxena,
  Nucl.\ Phys.\  B {\bf 710}, 425 (2005)
  [arXiv:hep-th/0406103];
 I.~Bena and N.~P.~Warner,
  Adv.\ Theor.\ Math.\ Phys.\  {\bf 9}, 667 (2005)
  [arXiv:hep-th/0408106];
  V.~Jejjala, O.~Madden, S.~F.~Ross and G.~Titchener,
  Phys.\ Rev.\  D {\bf 71}, 124030 (2005)
  [arXiv:hep-th/0504181].
  V.~Balasubramanian, E.~G.~Gimon and T.~S.~Levi,
  arXiv:hep-th/0606118;
  J.~de Boer, F.~Denef, S.~El-Showk, I.~Messamah and D.~Van den Bleeken,
  JHEP {\bf 0811}, 050 (2008)
  [arXiv:0802.2257 [hep-th]];
 S.~D.~Mathur,
  arXiv:0810.4525 [hep-th].

\bibitem{phase}
  S.~D.~Mathur,
  arXiv:0706.3884 [hep-th].

\bibitem{radiation}
V.~Cardoso, O.~J.~C.~Dias, J.~L.~Hovdebo and R.~C.~Myers,
  Phys.\ Rev.\  D {\bf 73}, 064031 (2006)
  [arXiv:hep-th/0512277];
B.~D.~Chowdhury and S.~D.~Mathur,
  Class.\ Quant.\ Grav.\  {\bf 25}, 135005 (2008)
  [arXiv:0711.4817 [hep-th]].


\bibitem{tunnel}
  S.~D.~Mathur,
  arXiv:0805.3716 [hep-th].
 
 \bibitem{time}
  S.~D.~Mathur,
  Int.\ J.\ Mod.\ Phys.\  D {\bf 18}, 2215 (2009)
  [arXiv:0905.4483 [hep-th]].

  \bibitem{scrambler}
 P.~Hayden and J.~Preskill,
  JHEP {\bf 0709}, 120 (2007)
  [arXiv:0708.4025 [hep-th]];
Y.~Sekino and L.~Susskind,
  JHEP {\bf 0810}, 065 (2008)
  [arXiv:0808.2096 [hep-th]].
  
  \bibitem{maldacena}
  J.~M.~Maldacena,
  Adv.\ Theor.\ Math.\ Phys.\  {\bf 2}, 231 (1998)
  [Int.\ J.\ Theor.\ Phys.\  {\bf 38}, 1113 (1999)]
  [arXiv:hep-th/9711200].


\bibitem{earlier}
M.~Parikh and F.~Wilczek,
  Phys.\ Rev.\  D {\bf 58}, 064011 (1998)
  [arXiv:gr-qc/9712077];
T.~Shimomura, T.~Okamura and T.~Mishima,
  Int.\ J.\ Mod.\ Phys.\  D {\bf 11}, 789 (2002);
  P.~Kovtun, D.~T.~Son and A.~O.~Starinets,
  JHEP {\bf 0310}, 064 (2003)
  [arXiv:hep-th/0309213];
V.~Cardoso, O.~J.~C.~Dias and L.~Gualtieri,
  Int.\ J.\ Mod.\ Phys.\  D {\bf 17}, 505 (2008)
  [arXiv:0705.2777 [hep-th]];
M.~Natsuume,
  Phys.\ Rev.\  D {\bf 78}, 066010 (2008)
  [arXiv:0807.1392 [hep-th]];
N.~Iqbal and H.~Liu,
  Phys.\ Rev.\  D {\bf 79}, 025023 (2009)
  [arXiv:0809.3808 [hep-th]];
  H.~Culetu,
  arXiv:0906.0920 [gr-qc];
C.~Eling and Y.~Oz,
  JHEP {\bf 1002}, 069 (2010)
  [arXiv:0906.4999 [hep-th]].



  
\end{thebibliography}
\end{document}